\documentclass{elsart3}
\usepackage{graphicx}
\usepackage{amssymb}
\usepackage{amsbsy}

\begin{document}
\begin{frontmatter}



\title{Linear polarization sensitivity of SeGA detectors}
\author[NSCL,MSU]{D. Miller\corauthref{DM}},
\ead{damiller@nscl.msu.edu}
\author[NSCL,MSU]{A. Chester},
\author[NSCL,MSU]{V. Moeller},
\author[NSCL,MSU]{K. Starosta},
\author[NSCL]{C. Vaman},
\author[NSCL]{D. Weisshaar}
\address[NSCL]{National Superconducting Cyclotron Laboratory, Michigan State University, East Lansing, MI 48824, USA}
\address[MSU]{Department of Physics and Astronomy, Michigan State University, East Lansing, MI 48824, USA}
\corauth[DM]{Corresponding Author: Tel.: +1-517-333-6416; fax: +1-517-333-5967.}
\begin{abstract}
Parity is a key observable in nuclear spectroscopy. Linear polarization measurements of $\gamma$-rays are a probe to access the parities of energy levels. Utilizing the segmentation of detectors in the Segmented Germanium Array (SeGA) at the NSCL and analyzing the positions of interaction therein allows the detectors to be used as Compton polarimeters. Unlike other segmented detectors, SeGA detectors are irradiated from the side to utilize the transversal segmentation for better Doppler corrections. Sensitivity in such an orientation has previously been untested. A linear polarization sensitivity $Q \approx 0.14$ has been measured in the 350-keV energy range for SeGA detectors using $\alpha$-$\gamma$ correlations from a \nuc{249}{Cf} source. 
\end{abstract}
\end{frontmatter}

\section{Introduction}
\subsection{Overview}
Identifying parity of energy levels in nuclei is vital in understanding phenomena far from the valley of beta-stability such as the emergence of new shell gaps and the migration of single-particle levels with unique parities \cite{Dob94}. When a nucleus decays from an excited state, the information about the parity of the state can be extracted from the linear polarization and angular distribution of the radiation. With knowledge of the angular distribution, determination of the parity change requires only the knowledge of the sign of the polarization.
\subsection{Compton Polarimeters}
Polarimeters based on the sensitivity of Compton scattering to polarization have been built and used for fifty years \cite{Fag59} by the nuclear structure community. The first and simplest setup consisted of two $\gamma$-ray detectors where one functioned as the scatterer and the second served as the analyzer. This setup has been optimized for maximum polarization sensitivity \cite{Wer95}  at the expense of photopeak efficiency using a small scatterer and four analyzing crystals. Composite detectors such as the Clover \cite{Duc99} are natural Compton polarimeters when analyzing scattering between neighboring crystals. The sensitivity is reduced due to the proximity of the analyzing region to the scattering point. However, the increase in the efficiency of these detectors offsets this decreased sensitivity.  The advent of segmented high-purity Ge detectors has presented a new tool for polarization determination. Like composite detectors, segmented detectors successfully confront the problem of maintaining sufficient efficiency while the small segment volume allows low-energy Compton events to be detected with good sensitivity. The polarization sensitivity in segmented detectors has been demonstrated \cite{Wer95,Sch94} as adequate for making such measurements.
\subsection{SeGA as a Compton Polarimeter}
\begin{figure}
\begin{center}
\includegraphics[width=18pc]{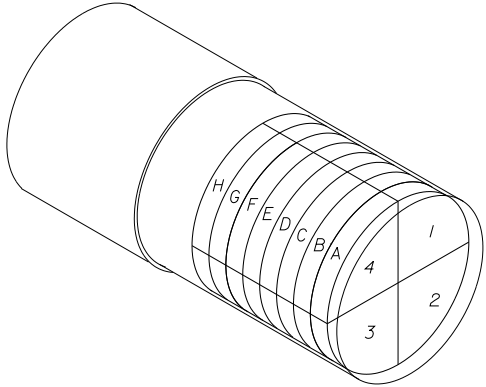}
\end{center}
\caption{The segmentation of a SeGA detector}
\label{fig:segmentation}
\end{figure}
\begin{figure}
\begin{center}
\includegraphics[width=18pc]{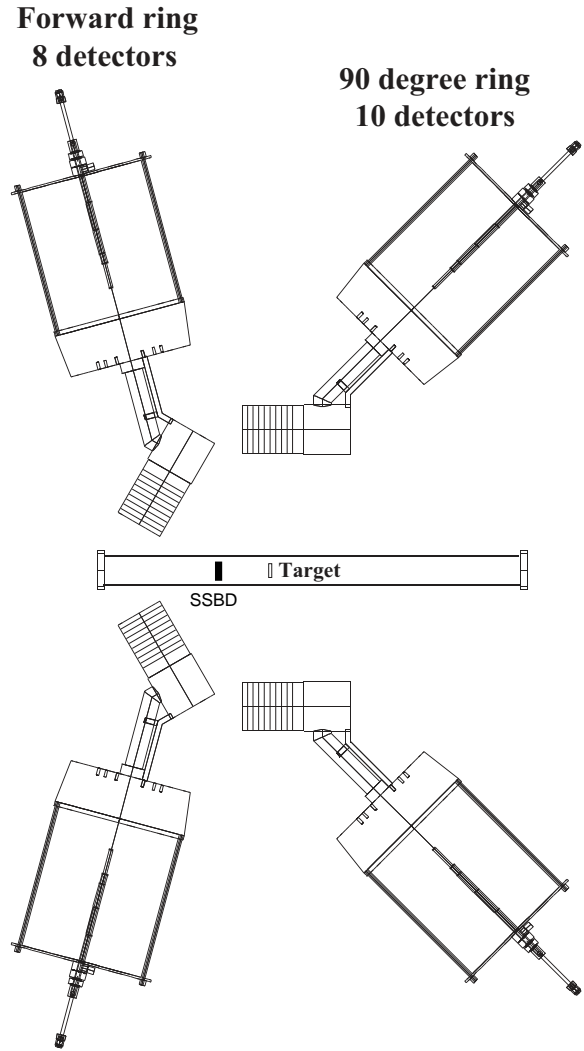}
\end{center}
\caption{A standard SeGA setup with an added silicon surface barrier detector used for this measurement}
\label{fig:segasetup}
\end{figure}

The Segmented Germanium Array (SeGA) \cite{Mue01} serves as the primary device for gamma-ray spectroscopy at the National Superconducting Cyclotron Laboratory (NSCL). The array consists of up to eighteen cylindrical detectors. Each detector is segmented into the four quadrants of the cylinder (similar to \cite{Wer95,Sch94}) and also into eight slices along its axis (Fig. \ref{fig:segmentation}) creating a total of thirty-two analyzing segments. SeGA detectors are used in configurations where they are irradiated from the side of the cylinder (Fig.~\ref{fig:segasetup}) as opposed to the face as in Refs. \cite{Wer95,Sch94}. In this scheme, the transversal segmentation provides a smaller opening angle to do better Doppler correction of a $\gamma$-ray's energy in fast beam experiments. The current study investigates the response of these detectors to polarized $\gamma$-radiation and the possibility of extending polarization measurements to exotic beams.
\section{Theoretical Framework}
If the axis of nuclear alignment and the photon's direction of propagation define the reaction plane, the theoretical linear polarization distribution of $\gamma$-rays can be written as $W(\theta,\xi)$ where $\theta$ is the angle between the quantization axis and the propagation direction and $\xi$ is the angle between the electric field of the $\gamma$-ray and the reaction plane\cite{Fag59} . From this, the polarization is defined as:
\begin{equation}
\label{poltoW}
P(\theta) = \frac{W(\theta,\xi=0) - W(\theta,\xi=90)}{W(\theta,\xi=0) + W(\theta,\xi=90)}.
\end{equation}
The quantity observed in experiment is the intensity of the energy peaks in the $\gamma$-ray spectrum corresponding to scattering of the photon at an angle $\phi$ within the detector. From these intensities, experimental asymmetry is defined as:
\begin{equation}
\label{eq:def_asymm}
A(\theta) = \frac{N(\theta,\phi=90) - N(\theta,\phi=0)}{N(\theta, \phi=90) + N(\theta,\phi=0)},
\end{equation}
where $N$ is the number of counts observed in the photopeak. The cross-section of Compton scattering is focused in the direction perpendicular to the electric field giving the relation $\phi = 90 - \xi$. For a polarimeter, the relation between the asymmetry and the polarization is given by:
\begin{equation}
\label{eq:poltoasym}
A(\theta) = Q(E_\gamma)P(\theta),
\end{equation}
where $Q(E_\gamma)$ is the polarization sensitivity of the detector. The sensitivity depends on the incident energy and is limited to that of a point-like polarimeter $Q_0$ as calculated from the Klein-Nishina cross-section:
\begin{equation}
\label{eq:def_q0}
Q_0(\alpha) = \frac{1+\alpha}{1+\alpha+\alpha^2} \; \mathrm{with} \; \alpha = \frac{E_\gamma}{m_e c^2}.
\end{equation}
For practical applications, the measured asymmetry $A$ must address geometric asymmetries present in the detector. These asymmetries are compared with those resulting from unpolarized $\gamma$-rays and characterized by the parameter $a$:
\begin{equation}
\label{eq:def_a}
a(E_\gamma) = \frac{N_{\mathrm{unpolarized}}(\theta,\phi=90)}{N_{\mathrm{unpolarized}}(\theta,\phi=0)}
\end{equation}
\begin{equation}
\label{eq:def_asymm2}
A(\theta) = \frac{N(\theta,\phi=90) - a N(\theta,\phi=0)}{N(\theta, \phi=90) + a N(\theta,\phi=0)}.
\end{equation}
To determine the sensitivity of detectors, the analysis is performed for $\gamma$-rays which have a well known polarization such as pure dipole transitions which have an angular distribution of
\begin{equation}
\label{eq:angdistro}
W(\theta) = 1 + a_2 P_2(\cos \theta).
\end{equation}
For pure dipole transitions, the polarization distribution reduces to the form \cite{Jon02}
\begin{equation}
\label{eq:poltheta}
P(\theta) = \pm\frac{3a_2 \sin^2 \theta}{2+2a_2-3a_2\sin^2\theta}
\end{equation}
where $a_2$ can be directly determined by investigating the angular distribution of $\gamma$-rays.

\section{Experimental Setup}
\begin{figure}
\begin{center}
\includegraphics[width=18pc]{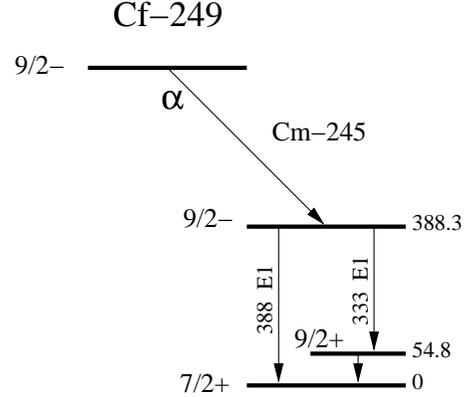}
\end{center}
\caption{Simplified decay scheme of $^{249}$Cf used to produce $\gamma$-rays of known polarization}
\label{fig:levels}
\end{figure}
\subsection{Source}
The experiment used $\alpha$-$\gamma$ coincidences from a \mbox{$\sim$ 1} $\mu$Ci \nuc{249}{Cf} source to provide polarized $\gamma$-rays\cite{Jon02,Fal70}. Following the $\alpha$ decay of $^{249}$Cf (see Fig.~\ref{fig:levels}), the daughter nucleus \nuc{245}{Cm} has two strong pure electric dipole transitions of 333 and 388 keV. These two $\gamma$-rays have polarizations that are perpendicular to each other which gives a difference in the sign of the polarization. Measuring polarizations of opposite signs simultaneously provides a control over systematic errors in the measurement.  
\subsection{Detector setup}
The $\alpha$-particles emitted from the source were detected by a 400 $\mu$m thick Si surface barrier detector located 11 cm from the source. The recoiling heavy ions were stopped in the 2-mil Pt foil backing of the source. Fifteen SeGA detectors were mounted in a standard frame which has two rings for detectors at $\theta=37^\circ$ and $\theta=90^\circ$ to the axis of $\alpha$ emission (See Fig. \ref{fig:segasetup}). The detectors are fixed at a radius of 24 cm to the source. Six detectors were located in the 37-degree ring, while nine were mounted in the 90-degree ring. The standard analog electronics for SeGA and the alpha detector were used to record the data. Gamma-ray singles were taken concurrently with the $\alpha$-$\gamma$ coincidences using a down-scaling factor of 125 which yielded singles spectra with roughly twice as many counts as the coincidence spectra. For the 333 keV transition, each detector recorded roughly 25,000 $\gamma$-singles as well as 7,000 coincidences over a time period of one hundred hours. The photopeak for the 388 keV transition contained four times as many counts.
\section{Analysis}
\subsection{Angular Distribution}
\begin{figure}
\begin{center}
\includegraphics[width=18pc]{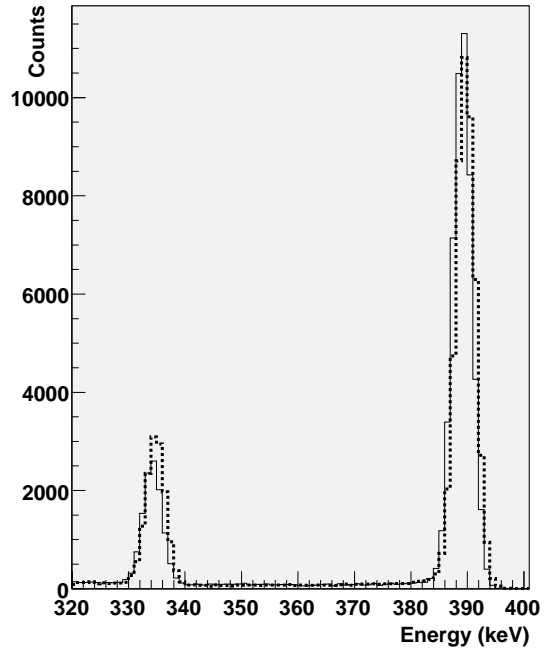}
\end{center}
\caption{Gamma spectrum for a detector in the 37$^{\circ}$ ring (solid) and the 90 $^{\circ}$ ring (dashed) showing the difference in angular distribution.}
\label{fig:angdistrospec}
\end{figure}

\begin{figure}
\begin{center}
\includegraphics[width=18pc]{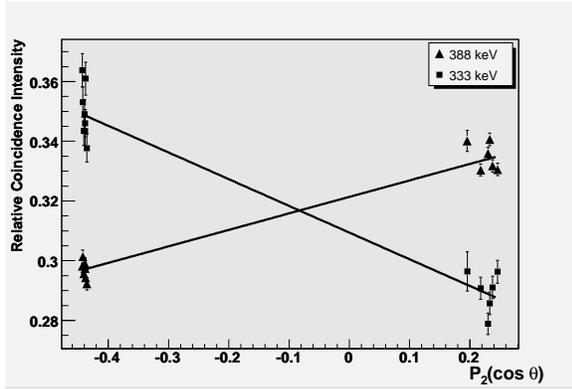}
\end{center}
\caption{The ratio of coincidences to single events as defined in Eq.~\ref{eq:doubtosing} observed in each of the detectors for the two $\alpha$-$\gamma$ transitions of interest with the line fit to determine the angular distribution coefficient $a_2$}
\label{fig:angdistro}
\end{figure}

The two electric dipole transitions of the daughter nucleus were examined for angular distribution. For pure dipole transitions, the polarization at a given angle is completely defined by the coefficients of the angular distribution from Eq.~\ref{eq:poltheta}, and the angular distribution of gamma rays has the form in Eq.~\ref{eq:angdistro}. The angular distribution can be observed in the spectrum for detectors at different angular positions in Figure~\ref{fig:angdistrospec}.
The number of $\alpha$-$\gamma$ coincidences observed in the experiment then is given by
\begin{equation}
\label{eq:doublesang}
N_d(\theta) = N_0 \epsilon_c(E_\alpha, E_\gamma, \theta) (1 + a_2 P_2(\cos \theta))
\end{equation}
where $N_0$ is the number of $\gamma$-rays emitted and $\epsilon_c$ is the coincidence efficiency for the experimental setup at a given combination of energies. This is directly related to the efficiency of the individual detectors
\begin{equation}
\label{eq:coinceff}
\epsilon_c(E_\alpha, E_\gamma) = \epsilon_\alpha(E_\alpha)\epsilon_\gamma(E_\gamma,\theta)
\end{equation}
The number of down-scaled (by a factor $f_{DS}$) singles is given by
\begin{equation}
\label{eq:singlesang}
N_s(\theta) = \frac{N_0 \epsilon_\gamma(E_\gamma,\theta)}{f_{DS}}
\end{equation}
The ratio of these is linear in $P_2$ taking into account the attenuation due to the extent of the detector
\begin{equation}
\label{eq:doubtosing}
\frac{N_d(\theta)}{N_s(\theta)} = f_{DS}\epsilon_\alpha(E_\alpha)W_{av}(\theta)
\end{equation}
where $W_{av}(\theta)$ is the average of the angular distribution function over the solid angle of the detector. This is independent of the individual germanium detectors' efficiencies. From the measured ratios defined in Eq.~\ref{eq:doubtosing}, the $a_2$ coefficients of 0.17 $\pm$ 0.01 for the 388 keV transition and -0.29 $\pm$ 0.02 for the 333 keV transition were extracted (Fig.~\ref{fig:angdistro}) which agrees with previous measurements done in the \nuc{249}{Cf} nucleus\cite{Fal70}.
\subsection{Geometric Asymmetry}
\begin{figure}
\begin{center}
\includegraphics[width=18pc]{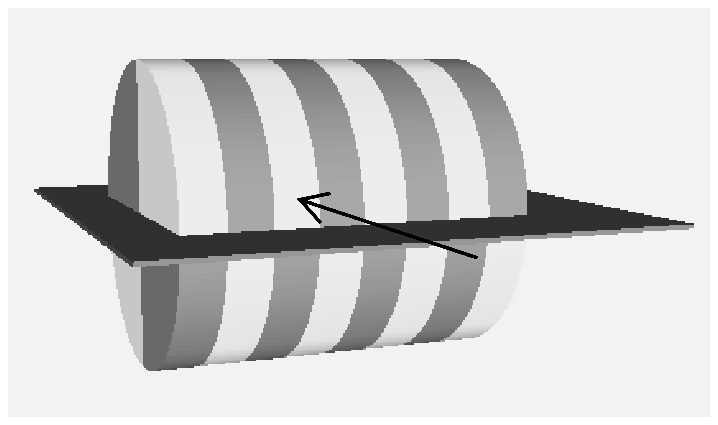}
\includegraphics[width=18pc]{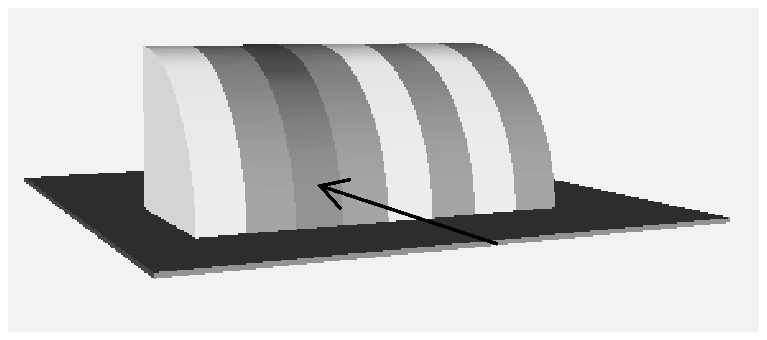}
\includegraphics[width=18pc]{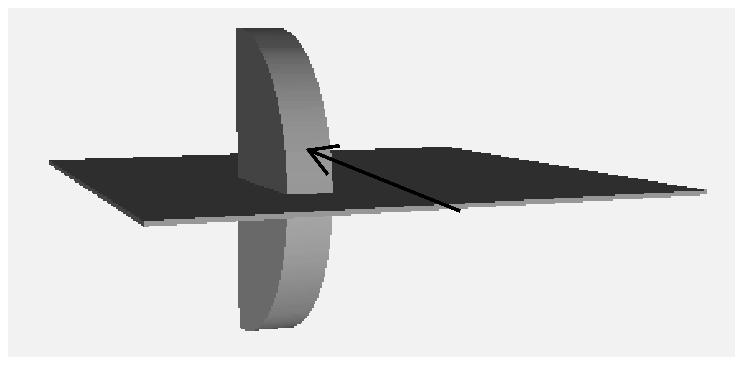}
\end{center}
\caption{(top) The geometry of a SeGA detector with respect to the reaction plane. $\gamma$-rays irradiate the detector from the side perpendicular to the symmetry axis and interact in a segment. (middle) Segments associated with parallel scattering for a given segment hit. (bottom) For the same segment, the associated perpendicular scattering}
\label{fig:scat}
\end{figure}
The scattering of the unpolarized $\gamma$-rays within the detector was examined for intrinsic asymmetries. The unpolarized $\gamma$-rays used were the decays from $^{245}$Cm where no $\alpha$ particles were detected, therefore no directional preference was present. Quantitatively, the ratio $a$ (Eq.~\ref{eq:def_a}) of scattering parallel and perpendicular to the reaction plane (Fig.~\ref{fig:scat}) was measured for each detector. Interactions were chosen in the given photopeak that only deposited energy in two segments corresponding to perpendicular or parallel scattering. The asymmetry of scattering in the two directions due to the detector geometry was considered separately for each of the dipole transitions. Over this energy range, the change in the asymmetry is negligible: $a(333) = 0.126 \pm 0.007$  and $a(388) = 0.128 \pm 0.007$ for the weighted averages of all the detectors. Much of the uncertainty is due to the individual differences in the detectors' geometric asymmetries; the standard deviation of $a$ for the sample of detectors is $\sigma=0.009$. It is also worthwhile to note that the selection of parallel (perpendicular) scattering events account for the 25\% (3\%) of the events registered in the detector in this energy range. These are the events which are considered to scatter by an angle of $90^\circ$ where the cross section is most sensitive to the polarization for low energies\cite{Fag59}.
\subsection{Sensitivity measurement}
\begin{figure}
\begin{center}
\includegraphics[width=18pc]{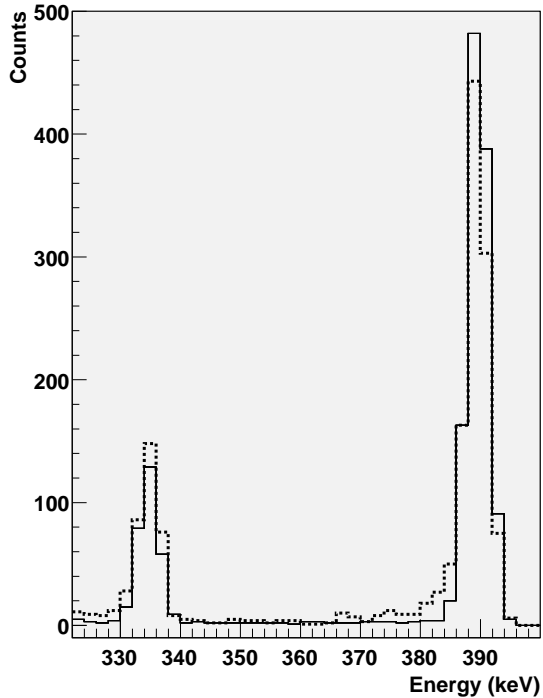}
\end{center}
\caption{Events scattering parallel (solid) and perpendicular (dashed) to the reaction plane for the two gamma transitions of interest in a detector renormalized to have the same number of counts in the total spectrum.}
\label{fig:paraperpcomp}
\end{figure}
\begin{figure}
\begin{center}
\includegraphics[width=18pc]{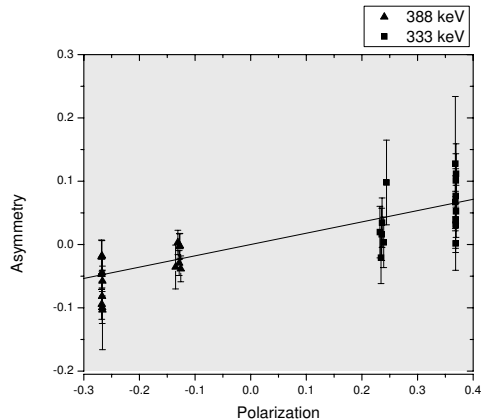}
\end{center}
\caption{Asymmetry observed in detectors for $\gamma$-rays of four different polarizations adjusted for the relative analyzing power of the Compton effect for the two energies.}
\label{fig:asymmVpol}
\end{figure}

Using each detector's geometric asymmetry from the singles measurement, the asymmetry $A$ (Eq.~\ref{eq:def_asymm2}) in the scattering was then observed for the $\alpha$-$\gamma$ coincidences. The finite angle span of the detector introduces a small uncertainty in the polarization of roughly 1\% which has been corrected by averaging the polarization over the solid angle of the detector. With this correction, the theoretical polarization for pure dipole transitions for the two $\gamma$-rays of interest at the two angles covered by SeGA (Eq.~\ref{eq:poltheta}) are -0.267 (7), -0.13 (2) ($E_\gamma = 388$), 0.24 (2), and 0.37 (2) ($E_\gamma = 333$) for the forward and backward rings respectively. 
\\ \indent
The measured asymmetries for each polarization group were compared to the analyzing power of the Compton effect, $Q_0$. The relative asymmetries observed for the respective groups were -0.08332 (7), -0.02242 (7), 0.04135 (3), and 0.09529 (2). The relative sensitivity $Q_{\mathrm{rel}} = Q/Q_0$ was then established from these asymmetries (Fig.~\ref{fig:asymmVpol}). The relative sensitivity was measured as 0.18 $\pm$ 0.02. 
\section{Discussion}
\subsection{Figure-of-merit}
Understanding the true effectiveness of the setup should encompass the time necessary to perform a sensitive measurement. Therefore, the effects of efficiency must be considered as well. The figure-of-merit (Eq.~\ref{eq:def_fom}) for a Compton polarimeter as defined in the work of Logan {\it et al.} \cite{Log73} takes into account the sensitivity and the coincidence efficiency (Eq.~\ref{eq:def_ec}) \cite{Sim83} of the detector. This value is inversely proportional to the amount of the time necessary to make a measurement of a desired precision with the detector.
\begin{equation}
\label{eq:def_fom}
FM = Q^2 \epsilon_c
\end{equation}
\begin{equation}
\label{eq:def_ec}
\epsilon_c(E_\gamma) = \frac{(N_\parallel + N_\perp) / 2}{N_{tot}}\epsilon_p\omega(E_\gamma)
\end{equation}
For a single SeGA detector in this energy range, the total photopeak efficiency $\epsilon_p\omega(E_\gamma)$ is approximately  $21.2 \cdot 10^{-4}$. Since 25\% and 3\% of the events go in the parallel and perpendicular direction respectively, the coincidence efficiency for polarization is $\epsilon_c \approx 2.99 \cdot 10^{-4}$ for a single detector. With this efficiency, the figure-of-merit is $FM \approx 5.9 \cdot 10^{-6}$.
\subsection{Comparison with other detectors}
\begin{table*}
\caption{Comparison of SeGA to other detector arrays for polarization measurements}
\begin{tabular*}{\textwidth}{@{\extracolsep{\fill}}l l l l l }
& $Q_{\mathrm{rel}}$\ & $Q$ & $\epsilon_c \cdot 10^4$ & $FM \cdot 10^6$ \\
\hline
POLALI \cite{Wer95}
 & 0.8 \it{2} & 0.5 \it{1} & 0.26 & 7.3  \\
Clover (EUROBALL)\cite{Duc99,Sta99}
 & 0.29 \it{2} & 0.23 \it{2} & 4.8 & 28 \\
Gammasphere \cite{Sch98,Dev96}
 & 0.08 \it{1} & 0.052 \it{7} & 18 & 4.8 \\
SeGA & 0.18 \it{2} & 0.14 \it{2} & 3.0 & 5.9 \\
\label{tab:fomcomparison}
\end{tabular*}
\end{table*}
The figure-of-merit for SeGA detectors is comparable to other detector setups within an order of magnitude. In Table \ref{tab:fomcomparison}, the key values for other arrays of detectors are listed. The sensitivity has been extrapolated to the low energy range ($\sim$ 350 keV) from the literature. Absolute photopeak efficiency has been estimated based on graphs available, and the figure-of-merit has been calculated according to Eq.~\ref{eq:def_fom} based on these estimates. Figure-of-merit values are considered to be accurate to 30\%. Values for an entire array are scaled from the individual detectors' figures-of-merit in a typical configuration. 
\\ \indent
Comparing this to the Clover results \cite{Duc99,Sta99}, the SeGA detectors each have a figure-of-merit that is about a fifth of the Clovers: roughly a third due to less sensitivity and a half due to less coincidence efficiency. However, this still results in a higher figure-of-merit than a standard 5-crystal Compton polarimeters (POLALI) due to the much larger efficiency afforded by the detector. However, each of these detector systems are well-suited for specific purposes such as SeGA's performance in fast beam $\gamma$-spectroscopy. Analysis to create similar quality of spectra in terms of signal-to-noise ratio can create an effective impact on the figure-of-merit.
\section{Outlook}
The SeGA detectors possess a figure-of-merit for polarization measurements that is comparable to other detectors of its kind \cite{Wer95,Sch94}. From the sensitivity measured, SeGA detectors should be able to measure polarization to an accuracy of $\Delta P \approx 0.3$ with 5000 counts in the photopeak for a given transition. Further investigation into the polarization sensitivity of SeGA detectors is warranted. In-beam sensitivity measurements of pure dipole and quadrupole transitions which have a high polarization would help complete the picture with the extension of sensitivity measurements up to higher energies. 
\\ \indent
Possessing the ability to do this with $\alpha$-$\gamma$ correlations gives the easiest method to repeat the measurement which is important if additional developments are tested to improved the detectors' sensitivities. Currently, most of the events which apply to polarization measurements are also allowed the largest range of scattering angles on which the sensitivity is dependent. One possible improvement involves digital data acquistion and $\gamma$-ray tracking. The ability to locate more accurately the first and second points of interaction in the detector would completely define the angle to the reaction plane as well as the scattering angle. 
\\ \indent
Polarization-sensitive detector arrays can explore extensive areas in a fragmentation facility, such as performing parity measurements in the ``island of inversion'' for odd-mass nuclei such as \nuc{31}{Mg} and \nuc{33}{Mg}. Identification of negative parity states in these nuclei will provide information about the evolving shell structure \cite{War90,Ots01}. Having the knowledge of the sensitivity of SeGA allows the feasibility of experiments to be considered providing the community with a useful tool to investigate nuclear structure.
\section*{Acknowledgments}
The work presented here was supported by the US National Science Foundation under the Grant PHY-0606007.

\end{document}